\documentstyle[aps,prd,epsfig]{revtex}
\newcommand{\bee}{\begin{equation}}
\newcommand{\ee}{\end{equation}}
\newcommand{\beea}{\begin{eqnarray}}
\newcommand{\eea}{\end{eqnarray}}

\newcommand{\half}{\frac{1}{2}}
\newcommand{\Tr}{\mbox{Tr}}

\begin{document}

\title{One loop matching coefficients for a variant overlap action--and
some of its simpler relatives}
\author{Thomas DeGrand}
\address{
Department of Physics,
University of Colorado, 
        Boulder, CO 80309 USA
}
\date{\today}
\maketitle
\begin{abstract}
I present one-loop perturbative calculations
of matching coefficients between matrix elements in continuum regulated 
QCD and lattice QCD with overlap fermions, with emphasis 
a recently-proposed variant discretization of the overlap.
These fermions have  extended (``fat link'') gauge connections. The scale
for evaluation of the running coupling constant (in the context
of the Lepage-Mackenzie fixing scheme) is also given.
 A variety of results (for additive mass renormalization,
local  currents, and some non-penguin four-fermion operators)
for naive, Wilson, clover, and overlap actions are shown.
\end{abstract}
%

%

\section{Introduction}
This paper presents one-loop perturbative calculations of matching coefficients
between matrix elements computed using a spatial lattice regulator
(and measured in lattice simulations) and their
equivalent continuum-regulated values. The principal lattice fermion action
studied is a recent formulation of overlap fermions
\cite{ref:neuberfer}) built from a ``kernel action'' with nearest and
next-nearest neighbor fermionic interactions and ``fat link'' gauge connections
\cite{ref:TOM_OVER},
specifically HYP-blocked \cite{Hasenfratz:2001hp} links.
 Along the way, I present some new results for a number
of simpler actions--naive, Wilson, and clover  fermions with fat links,
(including a few results for APE-blocked \cite{ref:APEblock} links), and
overlap actions with Wilson or clover action kernels.
I also present results for the scale at which the running coupling constant
is evaluated (the so-called $q^*$ value), using the scheme of
Lepage and Mackenzie\cite{ref:LM}.

Fat link actions were originally developed\cite{FATLINK}
 in order to improve the chiral properties of lattice fermions.
In the context of perturbation theory, 
  ``chiral improvement'' for nonchiral actions means having
small additive mass renormalization, vector and axial current renormalization
factors  nearly equal,  scalar and pseudoscalar
factors also nearly equal, and suppressed
 mixing of four-fermion operators between
different chiral sectors.  Fat link actions do all these things.
I have found that using these actions as kernels for overlap actions
seems to have computational advantages over using thin link actions.
 Overlap actions with fat links have
matching factors which are much closer to unity than overlap actions
built of thin link fermion action kernels.

 Several authors have published  calculations
of lattice-to-continuum matching factors for a variety of processes
for  the overlap action with the thin
 link Wilson action as its kernel
(for a partial list, see
 \cite{Ishibashi:1999ik,Alexandrou:2000kj,Capitani:2000da,Capitani:2000bm,Capitani:2001yq}). Perturbative calculations for more complicated actions
is in principle not much more difficult to do, once the Feynman rules
are constructed. One just has to organize the calculation, and not
carry it too far.
Intermediate results for matching coefficients (particularly Table
\ref{tab:fourf}) may be a more important part of the paper than
the tables of results.
These formulas are  certainly not new, but in most published papers they appear
embedded in action-specific discussions, and they are often hard to find until
after one has rederived them.

Whether or not perturbation theory can -- or should be -- used to
convert lattice numbers to continuum ones is  a matter of debate.
One sometimes sees blanket statements about the use of perturbation
theory in lattice calculations, as if perturbative
calculations per se were unaffected by the particular choice of lattice
action. But clearly different actions have different properties
when the cutoff is not taken away.
At nonzero cutoff, it is  a practical question for
any particular action, to ask how well perturbative calculations perform.
At sufficiently
small renormalized coupling the matching from one scheme to another
can be done with small uncertainty, the uncertainty being due to higher
order terms in a perturbative expansion in the renormalized coupling.

To be specific, let's write a one loop matching coefficient for an operator
which does not undergo mixing as
$Z= 1+\alpha_s/(4\pi)A$.
In today's simulations, typical values of $\alpha_s/(4\pi) \simeq 0.01-0.015$.
With standard actions, one often finds that the $A$'s are large.
For standard thin-link actions A is order 10-40, so the matching factor
differs from unity by an amount of 0.1 to 0.6.
In the actions considered here, typically the $A$'s are smaller
than about 5 in magnitude, corresponding to a shift from unity of 0.05-0.07.
One presumes, therefore, that the perturbative expansion of $Z$
is better behaved.
One's results are also less sensitive to the choice of scale $q$
for $\alpha_s(q^2)$ if the coefficient is small.

A brief report of perturbation theory for Wilson and clover actions
with APE-blocked links has been presented in Ref. 
\cite{Bernard:1999kc}. The qualitative features of the work done here
are all anticipated in that paper. Here
I will focus mainly on HYP blocking simply because it gives more
smoothing without delocalizing the action.
Applications to  the overlap in this work are also new.
Results for staggered fermions using HYP links have recently
 been presented by Sharpe and Lee\cite{ref:SharpeWlee}, and Lee
\cite{Lee:2002fj}
has discussed general features of perturbation theory for fat links.

In Sec. II, I describe the simple ingredients I used for lattice
perturbation theory.  Results of selected calculations for
currents are presented in Sec. III.
Appendices contain Feynman rules for the actions studied.

\section{Ingredients}

\subsection{Matching and Scale Setting}
A ``typical'' matching coefficient between a lattice regulated quantity
 and a dimensionally-regulated
quantity (modified minimal subtraction, $\overline{MS}$, for
example), for an operator which does not undergo
 mixing, is $Z= 1+(g^2/(16\pi^2)){\cal Z}$ where
\bee
{\cal Z} = I^F_{\overline{MS}} - I_{latt} .
\label{eq:Delta}
\ee
$I^F_{\overline{MS}}$ is the finite part of the continuum 
$2\omega=4-2\epsilon$ dimensional integral and $I^{latt}$ is a lattice
 integral. For a process with $N_f$ external fermion legs,
$I_i = \Gamma_i -  \Sigma_1 N_f/2$
where $\Gamma_i$ is the vertex renormalization and $\Sigma_1$ is the fermion
wave function renormalization. All these quantities will
be evaluated at one loop.
 In this work the
internal momentum integration variable will always label
 the momentum flowing through the gluon line.
I will always work at zero momentum for
external particles and fermion mass $m=0$, regulating any IR (infared)
 divergence of the diagram with a gluon mass $\lambda$.
 Of course, there is
nothing deep about these choices; they are made purely for expediency.
 $I_{\overline MS}$ will take the generic form
\bee
I_{\overline{MS}} = 16\pi^2\int {{d^{2\omega}k}\over{(2\pi)^{2\omega}}}
(\mu^2)^{2\omega}
{1 \over {k^2(k^2+\lambda^2)}}
(A + B\epsilon)
 = A \{ {1\over \epsilon} -
 \gamma_E + \log(4\pi) \} +A \log{{\mu^2}\over{\lambda^2}}+ A +B.
\label{eq:MSINT}
\ee
The term in curly brackets is simply discarded to give
$I^F_{\overline{MS}}$.
In the lattice integral, we can 
scale all dimensionful variables by appropriate powers of the lattice spacing
and write
\bee
I_{latt} = 16\pi^2\int_{ak}  I(ak,ap,am,a\lambda)
\ee
where $\int_{ak} = \prod_j \int_{-\pi}^{\pi} d (ak_j)/(2\pi)$ will be the
symbol for integration over the (rescaled) momentum hypercube.

If $I_{\overline{MS}}$
has an $A\log(\mu^2/ \lambda^2)$ term,  $I_{latt}$
will have an $A\log(1/(\lambda^2 a^2))$ IR divergence, too.
 It can be separated out by writing the integrand  as
\beea
I_{latt}  & = & 16\pi^2 \int_k (I(k,ap,am,a\lambda) -
 A{{ \theta(\pi^2 - k^2)}\over{k^2(k^2+a^2\lambda^2)}})
+ 16\pi^2\int {{d^4 k}\over{(2\pi)^4}}
 A{{ \theta(\pi^2 - k^2)}\over{k^2(k^2+a^2\lambda^2)}}
\nonumber \\
 & \equiv & J + A \log{{\pi^2}\over{a^2\lambda^2}}
\nonumber \\
\label{eq:LATSUB}
\eea
The first term of Eq. \ref{eq:LATSUB} is IR finite, and one can set
$\lambda=0$ in it. Thus
\bee
{\cal Z} = A \log( \mu^2a^2) +A(1-\log\pi^2) +B -J.
\ee

This would be the end of the story if we did not want to choose a scale
$q^*$ for
the coupling constant in Eq. \ref{eq:Delta}.
This is (unfortunately) a problem whose solution involves at least a two-loop
calculation (see the discussion in the Appendix of Ref. \cite{ref:GBS}).
Absent such a calculation, one could either let $q^*$ vary over
 some ``reasonable'' range ($1<q^*a<\pi$ at lattice spacing $a$, for example),
or construct some physically motivated ansatz for the scale.
The best known example of such a construction is that of
Lepage and Mackenzie\cite{ref:LM}: Imagine that one has some process
parameterized by a one loop integral, and assume that its higher
order behavior is dominated by gluonic vacuum polarization,
so 
\bee \alpha \int I(q) d^4q \rightarrow \alpha(q^*)\int I(q) d^4q  \approx
\int \alpha(q) I(q) d^4q .
\ee
 Expanding the coupling 
 $\alpha(q)= \alpha(q^*) - \beta_0 \alpha(q^*)^2 \log(q^2/q^{*2})+\dots$,
self consistency requires that the coefficient of $\alpha(q^*)^2$ vanish, or
\bee
\label{eq:LM}
\log(q^*) =
 {{\int d^4 q \log(q) I(q)}\over  {\int d^4 q I(q)}}
 \equiv {{\cal L}_1 \over {\cal Z}}
\equiv \langle\langle \log(q^2) \rangle\rangle
\ee
For the actions studied here, it often happens that the ${\cal Z}$ coefficient
of Eq. (\ref{eq:Delta}) is close to zero, and the calculation of $q^*$ using
Eq. (\ref{eq:LM}) produces absurd results. In that case, I substitute the
higher order expression of Hornbostel, Lepage and Morningstar \cite{ref:HLM},
\bee
\log(q^{*2}) = \langle\langle \log(q^2) \rangle\rangle \pm
 [-\sigma^2]^{1/2}
\label{eq:LM2}
\ee
with
\bee
\sigma^2 = \langle\langle \log^2(q^2) \rangle\rangle -
\langle\langle \log(q^2) \rangle\rangle^2
\ee
and
\bee
\langle\langle \log^2(q^2) \rangle\rangle
 = {{\int d^4 q\log^2(q^2) I(q)}\over{\int d^4 q I(q)}}
 \equiv {{{\cal L}_2} \over{\cal Z}}
\ee
is the weighted average analogous to Eq. (\ref{eq:LM}).

This is all well-defined for finite pure lattice expressions,
but when the operator has an anomalous dimension, it is not obvious what to do.
In that case, I evaluate Eq. \ref{eq:LM} following a prescription
 learned from C. Bernard \cite{Bernard:2002pc}. His proposal is to 
 construct a combination of dimensionally-regulated integrals
whose sum gives the term in curly brackets, and to subtract them from
the integral Eq. (\ref{eq:MSINT}) to produce a finite integral
in $4-2\epsilon$ dimensions. The $\epsilon \rightarrow 0$ limit of
the subtracted integral
can then be taken, leaving a UV-finite four-dimensional integral
for $\Gamma^F_{\overline{MS}}$
 which can be combined with
the integrand of Eq. \ref{eq:LATSUB}.
To do this, consider the two integrals
\bee
I_1 = 16\pi^2\int {{d^{2\omega}k}\over{(2\pi)^{2\omega}}}
(\mu^2)^{2\omega}
{1 \over {k^2(k^2+\mu^2)}}
(A + B\epsilon) = 
A \{ {1\over \epsilon} - \gamma_E + \log(4\pi) \} + A +B
\ee
and
\bee
I_2 = 16\pi^2\int {{d^{2\omega}k}\over{(2\pi)^{2\omega}}}
(\mu^2)^{2\omega}
{1 \over {(k^2+\mu^2)^2}}
(A + B\epsilon)
 = A \{ {1\over \epsilon} - \gamma_E + \log(4\pi) \} +B.
\ee
If $a+b=1$ and $a(A+B)+bB=0$, $aI_1 + bI_2$ is equal to the difference
$\Gamma_{\overline{MS}} - \Gamma^F_{\overline{MS}}$ and
we can combine the three expressions under one integral, then take
 the $\epsilon \rightarrow 0$ limit, to write
\bee
I^F_{\overline{MS}} = A J_1 + B J_2
\label{eq:finalnt}
\ee
where, pushing the IR divergence from the $\overline{MS}$ integral
into the lattice integral
\bee
J_1 = 16\pi^2\int {{d^{4}k}\over{(2\pi)^{4}}}
[ {{1-
  \theta(\pi^2 - k^2)}\over{k^2(k^2+\lambda^2)}} - {1 \over {(k^2+\mu^2)^2}} ]
\ee
and
\bee
J_2 = 16\pi^2\int {{d^{4}k}\over{(2\pi)^{4}}}
[ {1 \over {k^2(k^2+\mu^2)}} - {1 \over {(k^2+\mu^2)^2}}].
\ee
The  integrands of $J_1$, $J_2$, and $J$ are 
 then used in Eqs. \ref{eq:LM} or \ref{eq:LM2} (with appropriate
powers
of $\log(q^2)$, of course).  Naturally, the particular choice of integrals
$J_1$
and $J_2$ are not unique, but because these integrals are typically
small in magnitude this is not generally a practical problem.
Notice that the $q^*$ scale depends on $\mu$. I will present results
only for the case $\mu a = 1$.

Presumably other prescriptions can be devised. Their results will probably
only depend on the coefficients A and B in Eq. \ref{eq:MSINT}, so
I will tabulate those parameters below for the processes considered.

The definition of $q^*$ for matching coefficients for operators which
mix, like the electroweak penguin operators, can be made as follows:.
Eq. \ref{eq:LM} expands into a matrix equation
\bee
\log q^{*2} {\cal Z} = {\cal L}_1
\label{eq:matLM}
\ee
where now ${\cal Z}$ and ${\cal L}_1$ are matrices.
There is a basis in which $\log q^{*2}$ is diagonal, found
by solving the eigenvalue equation
 ${\cal L}_1 {\cal Z}^{-1} - 1\log q^{*2}=0$,
 and in the original operator basis, the matrix of scales is
$\log q^{*2}= W (\log q^{*2})_{diag} W^{-1}$, where $W$ is the
basis-transformation matrix. Each eigenvector's scale
can be translated into a coupling, and
\bee
Z_{ij} = \delta_{ij} + W [\alpha_s(q^*)/(4\pi)]_{diag} W^{-1}{\cal Z}.
\ee
The higher order formula Eq. \ref{eq:LM2} does not have an obvious matrix
transformation. One possibility is to rotate the higher order
matrix ${\cal L}_2$ into the basis which diagonalizes 
${\cal L}_1 {\cal Z}^{-1}$, and use the diagonal entries in 
Eq. \ref{eq:LM2}.

This is unsatisfactory, but again, the small size of ${\cal Z}$
and the ${\cal L}_i$'s for fat link actions means that the
 spread in the $Z-$factor 
remains small as the coupling is varied.

\subsection{Actions and Feynman rules}
Feynman rules for ordinary (non-overlap) discretizations of the
Dirac operator can be constructed using standard techniques.
I will parameterize the free massive Dirac operator as
\bee
d(p,m) = i\gamma_\mu\rho_\mu(p) + \lambda(p) + m;
\label{eq:dirac}
\ee
I implicitly assume the normalization that at
 small $p$, $\rho_\mu(p)\simeq p_\mu$ and $\lambda(p)=O(p^2)$.
(This factorization might not be appropriate for an approximate fixed-point
action, designed to follow out a renormalized trajectory as the mass is varied:
$\rho_\mu$ and $\lambda$ would both be functions of the bare mass
\cite{DeGrand:1999gp}.)
Of course, the propagator is the inverse of $d$.

 While I will not specify an explicit form for the gluon propagator in 
any expression,
I will only present results for
the Wilson gauge action. Its propagator includes
an IR regulator mass $\lambda$ and gauge parameter $\xi$
\bee
G_{\mu\nu} = {{\delta_{\mu\nu} + (\xi -1) \hat k_\mu \hat k_\nu/\hat k^2}
\over {\hat k^2} +\lambda^2}
\ee
with $\hat k_\mu = 2/a \sin(k_\mu a/2)$ and 
$\hat k^2 = \sum_\mu \hat k_\mu^2 $. Propagators for other actions
 can be constructed
by inverting the gauge field's linearized equation of motion.
It is convenient to be able to vary the choice of gauge to test results.

The fermion and gluon
propagators are of course diagonal in color space.

The massless overlap operator is defined so that its eigenvalues
 lie on a circle of radius $x_0$, so
\bee D(0) = x_0(1+ {z \over{\sqrt{z^\dagger z}}} )
\label{eq:gw}
\ee
where $z = d(-x_0)/x_0 =(d-x_0)/x_0$ and $d(m)=d+m$ is the massive
 Dirac operator for mass $m$ 
 (i.e. $x_0$ is equivalent to a negative mass term
 and $d= i \gamma_\mu\rho_\mu +\lambda$ as above.)
The overall multiplicative factor of $x_0$ is a useful convention; when
 the Dirac operator $d$ is thought of as ``small'' and Eq. \ref{eq:gw} is
expanded for small $d$, $D \simeq d$.
Feynman rules for the Wilson overlap action have been given by Ref. 
\cite{Ishibashi:1999ik}, and can be straightforwardly be adapted for any kernel
action.

For overlap actions, it is customary to define the massive overlap
operator in terms of the massless one. as
\bee
D(m_q) = ({1-{m_q \over{2x_0}}})D(0) + m_q.
\ee
This results in an annoying entanglement of the mass with the vertices,
complicating a direct computation of the running mass.
Fortunately, we can compute  the multiplicative renormalization
factor for the fermion mass indirectly as the inverse
 of the scalar current renormalization factor, and we can evaluate the
latter expression at zero quark mass.

\begin{figure}[thb]
\begin{center}
\epsfxsize=0.6 \hsize
\epsffile{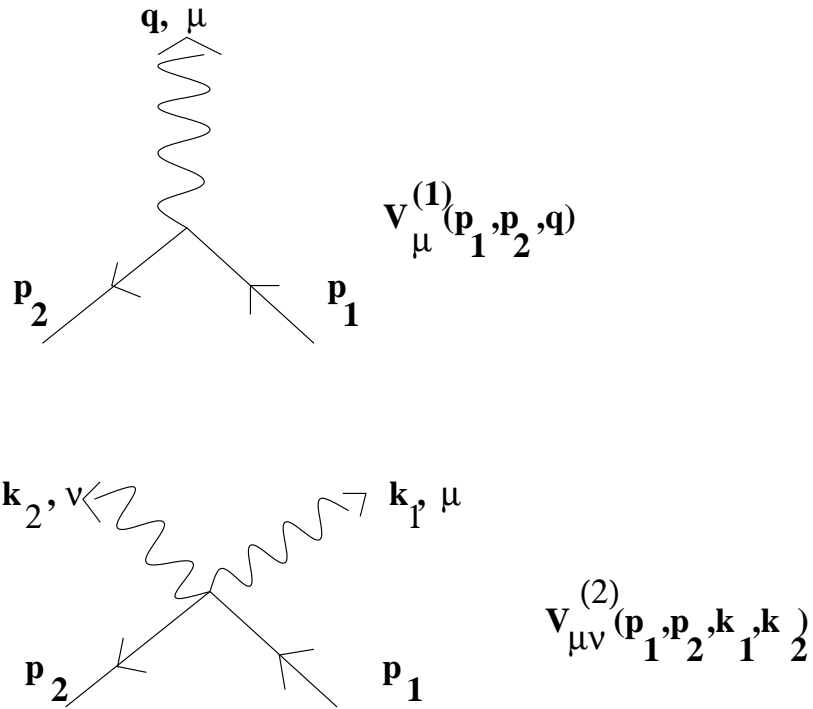}
\end{center}
\caption{
Three-point and four-point vertices, showing my convention for momentum
flows.}
\label{fig:vertices}
\end{figure}

\subsection{Unitary Fat Links}
We will be concerned only with unitary fat links, gauge connections
which are themselves elements of the gauge group, even though they may be
built of sums of products of the original thin links of the simulation.
For smooth fields the fat links
have an expansion $V_\mu(x) = 1 + i a B_\mu(x)+\dots$
and the original thin links have an expansion 
$U_\mu(x) = 1 + i a A_\mu(x)+\dots$.
For computations of
2- and 4-quark operator renormalization/matching
constants at one loop, only the linear part of the
relation between fat and thin links is needed,
 and it can be parameterized as
\bee
\label{eq:linA}
B_\mu(x) = \sum_{y,\nu} h_{\mu\nu}(y) A_\nu(x+y)\ .
\ee
Quadratic terms in (\ref{eq:linA}), which would only be relevant for tadpole
graphs, appear as commutators and therefore do not contribute, since
tadpoles are symmetric in the two gluons
\cite{Patel:1992vu,Bernard:1999kc,ref:SharpeWlee,Lee:2002fj}.
In momentum space, the convolution of Eq.~(\ref{eq:linA}) becomes 
 a form factor
\bee
\label{eq:momspA}
 B_\mu(q) = \sum_{\nu} \tilde h_{\mu\nu}(q) A_\nu(q)\ .
\label{eq:linkBA}
\ee
The reader could think of fat-link action Feynman rules as being constructed
in two levels: First find the vertices for actions with ordinary thin 
links, and
then replace the thin link by a unitary fat link.
Each quark-gluon vertex gets a form factor 
$h_{\mu\nu}(q)$, where $q$ is the gluon momentum.  If all gluon lines
start and end on fermion lines, then, effectively, the
gluon propagator changes into
$
G_{\mu\nu} \longrightarrow  \tilde h_{\mu\lambda}G_{\lambda\sigma}
\tilde h_{\sigma\nu}$.

(Notice that this is a perturbative realization of the statement that
fat link fermion connections can be converted into thin link fermion
connections by redefining the fat link variable as an ordinary thin link
variable, but with a more complicated pure gauge action.)

\begin{figure}[!thb]
\begin{center}
\epsfxsize=0.7 \hsize
\epsffile{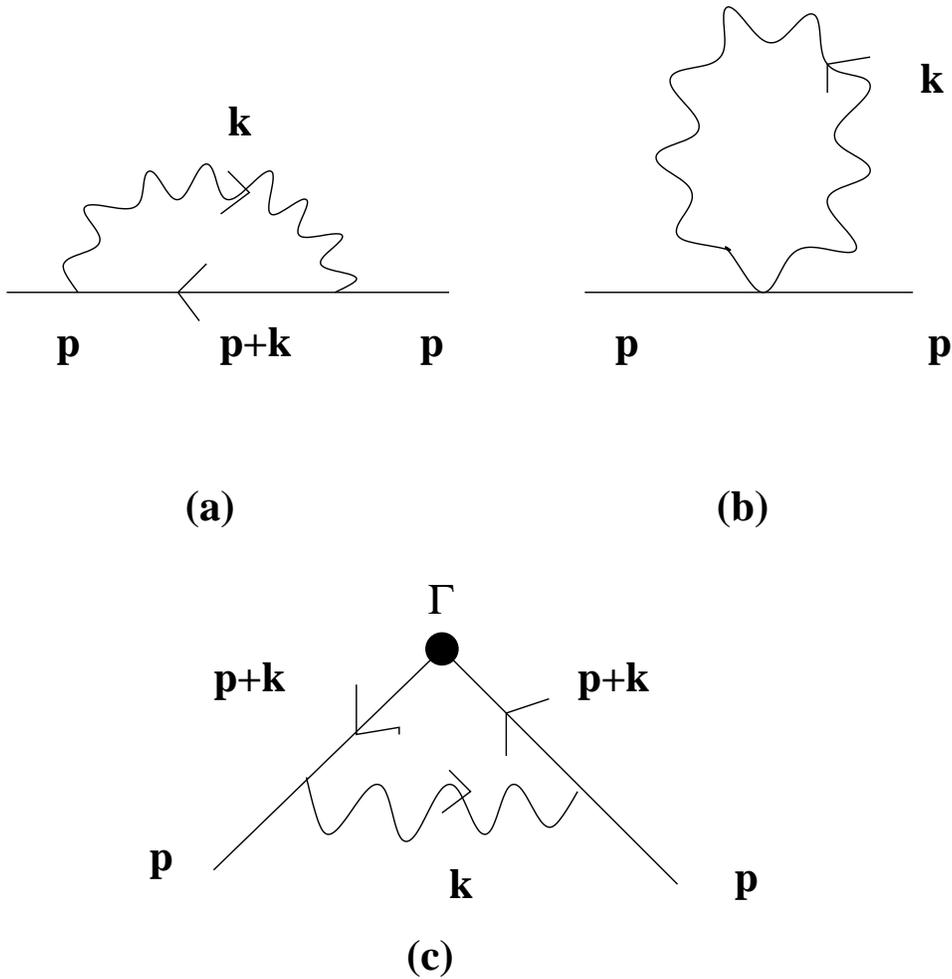}
\end{center}
\caption{
The three one-loop diagrams (a) ``sunset'' and (b) ``tadpole'' fermion
self energy and (c) vertex renormalization, showing my convention for momentum
flows.
 }
\label{fig:oneloop}
\end{figure}

Form factors for APE and HYP-blocked links are given in Appendix B.

\subsection{One-loop Diagrams}

The ``standard method'' for carrying out a lattice perturbation theory
calculation is to construct the integrand $I(k,p,m,\lambda,a)$
 analytically as a combination of terms multiplying Dirac matrices,
and then to project out the desired Dirac structure,
producing a single scalar expression $\hat I(k,p,m,\lambda,a)$.
Possible divergent terms are subtracted out, and the
 integral is performed using a Monte Carlo routine like
VEGAS\cite{Lepage:1980dq}.
The lattice actions I have studied are rather complicated (as are most improved
actions) and this procedure appears at first sight to be somewhat daunting.
The difficult part of the calculation is the Dirac reduction, particularly
as the action (and its Feynman rules) become complicated.
One can, of course, do the Dirac algebra using a symbolic manipulation
code. However,  there is a simpler path for the typical lattice practitioner:
 Take the parts of the programs
which one has already written to do full-scale numerical simulations
of the lattice action, and extract the routines which multiply
Dirac matrices times spinors.
Write the vertices and propagators as explicit
$4\times 4$ matrices, string the expressions together, a
 and let
the computer do all the Dirac multiplication and projection
as if it were doing a standard lattice Monte Carlo calculation.
All calculations factorize into a set of routines for each action of
interest, and a set of integrands (with appropriate projection algorithms)
for each coefficient.

This ``method'' is computationally inefficient, but it is easy to
study a wide variety of lattice actions.

I will parameterize the orientation of fermion and gluon momenta
through the vertices as shown in Fig. \ref{fig:vertices}.

The fermion self energy is parameterized for small fermion momentum $p$ and 
small mass $m$ as
\bee
\Sigma(p,m) = \Sigma_0 + i \gamma \cdot p \Sigma_1 + m \Sigma_2.
\ee
If nonzero, $\Sigma_0$ is (minus) the additive mass renormalization.
$\Sigma_1$ is the wave-function renormalization, needed for
all external lines in vertex functions. The translation of
the quark mass from lattice to $\overline{MS}$ regularization ($Z_m)$ 
is proportional to the difference
$\Sigma_1 - \Sigma_2$. While $\Sigma_0$ is finite (in lattice units;
it is proportional to $1/a$ and so diverges in the continuum limit),
 $\Sigma_1$ and $\Sigma_2$ are infared divergent.

$\Sigma$ itself is a sum of two terms, as shown in Fig. \ref{fig:oneloop}.
The ``sunset'' graph, Fig. \ref{fig:oneloop}(a), uses first order
vertices:
\bee
\Sigma_a = g^2 \int_k
 V^{(1)}_\mu(p,p+k,-k) S(p+k)  V^{(1)}_\nu(p+k,p,k)
G_{\mu\nu}(k).
\ee
The ``tadpole'' graph, Fig. \ref{fig:oneloop}(b), is
\bee
\Sigma_b = -{{g^2}\over 2} \int_k
 V^{(2)}_{\mu\nu}(p,p,k,-k)
G_{\mu\nu}(k).
\ee
Momenta are labeled as shown in the figure. Recall that we usually set $p=0$.

To extract $\Sigma_1$ and $\Sigma_2$, I expand the propagators
and vertices in a power series in $p$ and $m$ (respectively) and keep
the leading term. If the free Dirac operator is of the form of Eq. 
(\ref{eq:dirac}) this is straightforward to do. I have performed this expansion
analytically, since these two quantities are IR divergent, and
require subtraction. Below, the quantity $S_1$ will be used, with
where $\Sigma_1 = g^2C_F/(16\pi^2)S_1$.

The renormalization of currents involves both the vertex graph
and $\Sigma_1$. I compute the vertex graph simply by taking
the amplitude
\bee
V^{\Gamma} = \int_k  V^{(1)}_\mu(p,p+k,-k)] S(p+k) \Gamma
[S(p+k) V^{(1)}_\nu(p+k,p,k)]
G_{\mu\nu}(k),
\ee
evaluating it at $p=0$, $m=0$ (with gluon mass $\lambda$)
and tracing it with the appropriate Dirac projector, before doing
the integral. For the vector and axial currents, I average over traces
in the four cardinal dimensions.

An alternate parameterization of the vertex allows a connection to
matching coefficients of the four
 fermion operators of the effective field theory of
 electroweak interactions: Write
\bee
V^\Gamma =  K_0 \Gamma  +
 K_1 \gamma_\mu \Gamma\gamma_\mu  
+K_2 \gamma_\mu \gamma_\nu \Gamma \gamma_\nu \gamma_\mu 
+ \dots.
\label{eq:vadesire}
\ee
 Grouping propagator-vertex products,
\bee
V^\Gamma = \int_k  T^1(p,k) \Gamma T^2(p,k)
G_{\mu\nu}(k), 
\ee
with  $T^1(p,k) =  V^{(1)}_\mu(p,p+k,-k)S(p+k)$
and $T^2(p,k) = S(p+k) V^{(1)}_\mu(p,p+k,-k)$. One can find the $K$'s
by projecting the $T$'s onto elements of the Clifford algebra,
$T = T_0 + \gamma_\mu T_1 + \sigma_{\mu\nu}T_2 +\dots$, with
$T_0=1/4 \Tr\ T$, $T_1=1/4 \Tr \ \gamma_\mu T$, and
\bee
T_2 = 1/8 \sum_{\mu \ne \nu}\Tr (\gamma_\mu \gamma_\nu - \gamma_\nu \gamma_\mu)T.
\ee
Direct computation plus a consideration of lattice symmetries then
allows us to extract the separate terms of Eq. (\ref{eq:vadesire}), 
(here in Feynamn gauge) as
\bee
K_0= \int_k(  T^1_0 T^2_0 - 2 T^1_2 T^2_2)G_{\mu\mu}.
\ee
and
\bee
K_2= \int_k \half T^1_2 T^2_2 G_{\mu\mu}.
\ee

The Wilson and clover actions have only the listed terms ($K_0$, $K_1$,
 $K_2$).
 The overlap action only has nonzero $K_0$ and $K_2$ terms.
The continuum calculation with massless fermions only has nonzero $K_2$.
Finally, actions which only approximate an overlap action could in
principle span the Clifford algebra, although the coefficients
of the other terms would be small if the action were a good approximation.
This happens for the non-overlap planar action with HYP links.

The $K_1$ term makes its presence felt most malignantly in the
one loop correction to four fermion operators, where it
is responsible for  ``bad'' operator mixing into opposite-chirality
operators. It poisons lattice calculations of $B_K$ with
Wilson-type quarks.

To find the full Z-factor, and $q^*$, we also need the
coefficients of $J_1$ and $J_2$
(in the notation of Eq. \ref{eq:finalnt}). These are recorded in Table
 \ref{tab:vec}.
These results are certainly not new, but it is useful to collect them.

If only $K_0$, $K_1$, and $K_2$ are nonzero, we can immediately
write down relations (which appear many times in the literature)
between the lattice parts of the matching coefficients of the
scalar (S), pseudoscalar (P), vector (V), axial vector (A), and tensor (T)
currents: The only one we will need below is
\bee
I_S - I_P = 2(I_V - I_A) = 8K_1.
\label{eq:id3}
\ee
These relations can be used to relate the matching factors for four-fermion
operators to those for bilinears.  The earliest
reference I can find for this decomposition is by
 Martinelli \cite{Martinelli:1983ac}
and it  has been written down most usefully
by Gupta, Bhattacharya, and Sharpe \cite{ref:GBS}. A complication which 
arises in this case is the prescription used to define $\gamma_5$
away from four dimensions. The combination of $1/\epsilon$ factors
from integrals related to Eq. \ref{eq:MSINT} and $\epsilon$ factors
from the Dirac algebra is different in the bilinear and four-fermion
cases, meaning that a particular four-fermion $Z-$factor
into a particular continuum convention  is a linear
combinations of bilinear $Z_i$'s plus extra constant terms.

I found it most straightforward way to find these constants was to
 do the continuum Dirac algebra using the techniques of Ref. \cite{ref:greek}
(basically copying the examples of Ref. \cite{ref:buras98}).
In order to extract the momentum scale $q^*$, we need to separate the $A$ and
$B$ coefficients of Eq. \ref{eq:MSINT}.  Most lattice calculations do not
include penguin graphs, and four
operators are needed for the
most frequently performed four fermion matrix elements, combinations of
\bee
O = ( \bar q^{(1)}_\alpha \Gamma_1  q^{(2)}_\beta) \otimes
( \bar q^{(3)}_\gamma \Gamma_2 \hat  q^{(4)}_\delta).
\ee
Special cases are (a) $\Gamma_1=\Gamma_2 = \gamma_\mu(1-\gamma_5)$:
if color labels $\alpha=\delta$, $\beta=\gamma$, $O=O_1$;
if  $\alpha=\beta$, $\gamma=\delta$, $O=O_2$;
and  (b)the isospin 3/2 operators for electroweak penguins,
\bee
O_7^{3/2}=(\bar s_\alpha \gamma_\mu(1-\gamma_5) d_\alpha)
[(\bar u_\beta \gamma_\mu(1+\gamma_5) u_\beta)
-(\bar d_\beta \gamma_\mu(1+\gamma_5) d_\beta)]
+ (\bar s_\alpha \gamma_\mu(1-\gamma_5) u_\alpha)
(\bar u_\beta \gamma_\mu(1+\gamma_5) d_\beta)
\ee
and
\bee
O_8^{3/2}=(\bar s_\alpha \gamma_\mu(1-\gamma_5) d_\beta)
[(\bar u_\beta \gamma_\mu(1+\gamma_5) u_\alpha)
-(\bar d_\beta \gamma_\mu(1+\gamma_5) d_\alpha)]
+ (\bar s_\alpha \gamma_\mu(1-\gamma_5) u_\beta)
(\bar u_\beta \gamma_\mu(1+\gamma_5) d_\alpha)   .
\ee
 Ingredients for the non-penguin
mixing factors are given in Table \ref{tab:fourf}.

\section{Some Examples}
I have studied a large variety of lattice actions with thin and fat links.
Numerical integrals are checked, when possible, by comparison against
published results. For currents, I vary the gauge choice $\xi$ and check that
integrals (and integrands) remain gauge invariant. As a general
rule using double precision insures that the integrands
 are gauge invariant point by point 
 to a few parts in $10^{5}$.

My results for any standard thin link
action (Wilson, clover, Wilson overlap, \dots)
are (with one exception) not new, and there is no point in republishing old
 results already
in the literature. Results for a standard action with
HYP links are mostly unpublished, so I will show them in tables.
I will show results for the clover and planar actions, since
the HYP clover action might be an attractive action for simulations.
I will also tabulate results for the planar action, with thin
and HYP-blocked links, and the HYP-blocked planar overlap.

HYP blocking is characterized by three parameters
 with  ``preferred'' values $\alpha_1=0.75$, $\alpha_2=0.6$, $\alpha_3=0.3$.
While HYP blocking is typically presented (and used) at this
specific value of these coefficients, it is very useful
to show how the matching coefficients vary with the degree of fattening
of the link. I will do this simply by multiplying the standard
HYP coefficients by an overall scale factor, and tune the scale
factor from zero (corresponding to a thin link) to 1 to 1.5.
There are many other ways to tune HYP blocking, of course.

 With HYP smearing, the first-order
formula Eq. (\ref{eq:LM}) is small because
of cancellations of negative and positive contributions in the integral,
and the second order formula is often needed. The reader will note
 many ``cusps'' in the $q^*$ plots for HYP actions as I switch from first
to second order $q^*$'s.
 In contrast, as a general rule, under increased
APE-smearing $q^*$ usually falls slowly towards zero, and the first-order
formula for $q^*$ works well, unless the actual matrix element
vanishes.

\subsection{Additive mass renormalization}
\begin{figure}[thb]
\begin{center}
\epsfxsize=0.7 \hsize
\epsffile{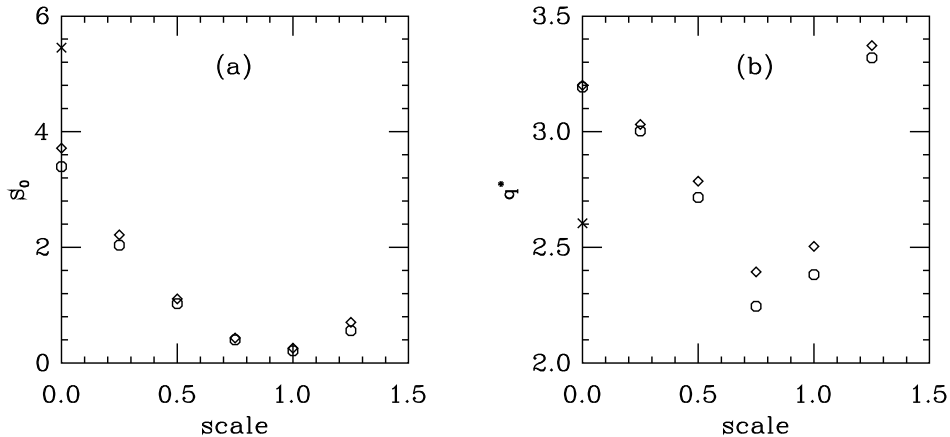}
\end{center}
\caption{
(a) $S_0$  parameterizing additive mass renormalization for thin link
 Wilson (cross), and HYP-link link clover (diamond)
and planar  (octagons) fermions and (b) their momentum scale $q^*$
with  HYP blocking, with the optimum parameters scaled by the shown overall 
scale factor.
}
\label{fig:sigmalog}
\end{figure}

All of the features of fat link perturbation theory can be seen in the
additive mass renormalization for non-overlap actions.
Fig. \ref{fig:sigmalog} shows the additive mass renormalization for
thin link Wilson and clover fermions, with $c_{SW}=1$, and for
the planar action with HYP-blocked
links. The graphs show $S_0$, with the definition
 $\delta m = \alpha_s(q^*) S_0$. 
We see that all these
thin link actions have large additive mass renormalization.
The addition of the clover term reduces
$S_0$ by about half, but it is still big. Smearing the
gauge fields has a dramatic effect on $S_0$, until the
scale factor for HYP blocking exceeds unity. At this point the
blocking enhances the large gluon momentum region in the integrand rather than
suppressing it. This effect shows little dependence on the choice of
fermion action
(planar or clover).

\subsection{Fermion Bilinears}

Next we turn to results for fermion bilinears, parameterized
as $Z_i = 1 + z_i g^2(q^*)C_F /(16\pi^2) $.
Table \ref{tab:hyppl} shows Z-factors for currents for
the thin link and HYP-blocked
planar action. The values quoted in all tables have an uncertainty
smaller than $\pm 1$ in their rightmost digit.

Table \ref{tab:naive} shows Z-factors for the thin link and HYP-blocked
naive fermions. These results form a (tiny) subset of an extensive
calculation of matching factors for staggered fermions by Sharpe and W. Lee
\cite{ref:SharpeWlee}. The $q^*$ values are new.

\begin{figure}[thb]
\begin{center}
\epsfxsize=0.7 \hsize
\epsffile{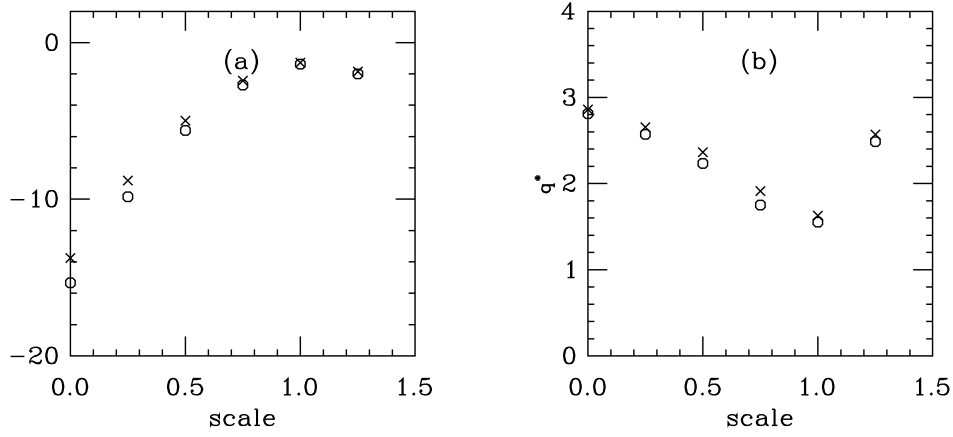}
\end{center}
\caption{
(a) $z_V$ and $z_A$, the coefficients of $g^2C_f/(16\pi^2)$ for the local
vector (octagons) and axial current renormalization constants (crosses)
 for clover fermions
with HYP blocking.
 (b) Momentum scale $q^*$ for the local vector and axial currents.
}
\label{fig:zvzacl}
\end{figure}

\begin{figure}[thb]
\begin{center}
\epsfxsize=0.7 \hsize
\epsffile{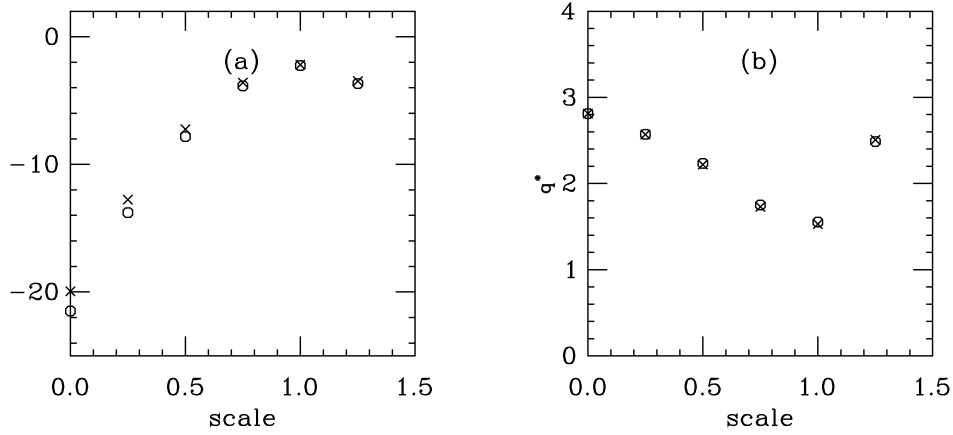}
\end{center}
\caption{
(a) $z_V$ and $z_A$ and their $q^*$'s, for the HYP-planar action,
labelled as in Fig. \protect{\ref{fig:zvzacl}}.
}
\label{fig:zvzapl}
\end{figure}

\begin{figure}[thb]
\begin{center}
\epsfxsize=0.7 \hsize
\epsffile{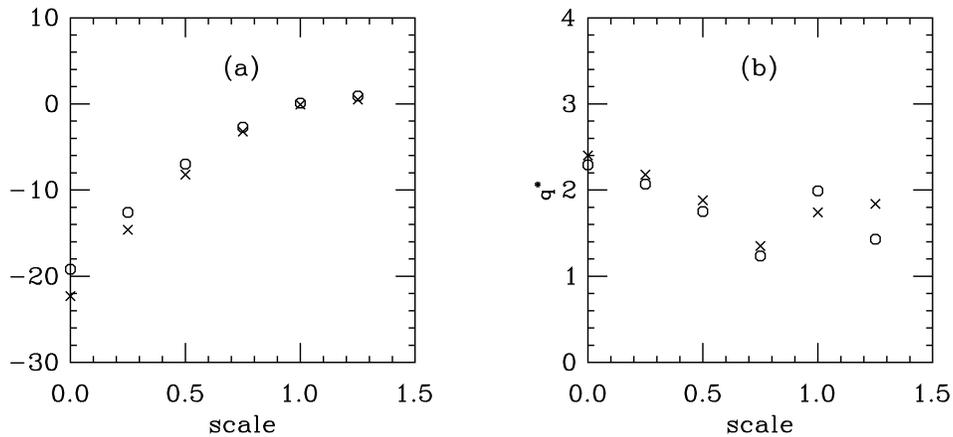}
\end{center}
\caption{
(a) $z_S$ and $z_P$, the coefficients of $g^2C_F/(16\pi^2)$ for the local
scalar and pseudoscalar current renormalization constants (evaluated
at $\mu a = 1$)
 for clover fermions, labelled by octagons and crosses, respectively.
 (b) Momentum scale $q^*$.
}
\label{fig:zpscl}
\end{figure}

\begin{figure}[thb]
\begin{center}
\epsfxsize=0.7 \hsize
\epsffile{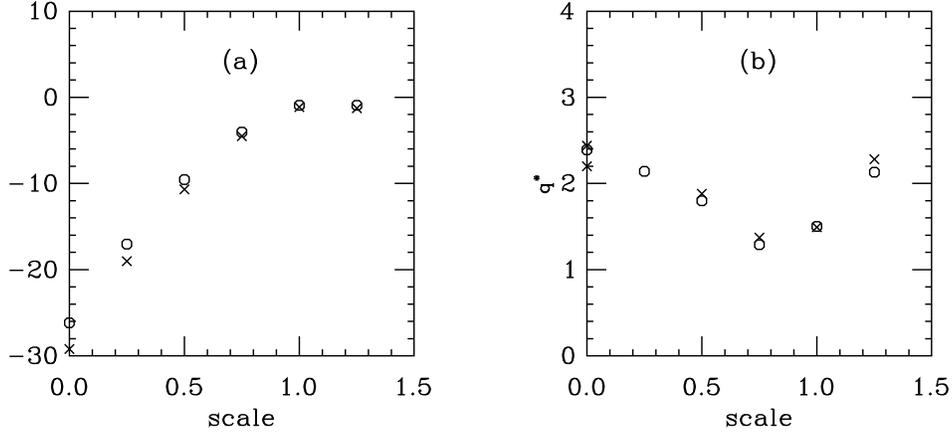}
\end{center}
\caption{
(a) $z_S$ and $z_P$ and their $q^*$'s, for HYP-planar action,
labelled as in Fig. \protect{\ref{fig:zpscl}}.
}
\label{fig:zpspl}
\end{figure}

Matching factors and $q^*$ scales for the local vector and axial vector
currents  for clover fermions and planar fermions as a function of fattening
strength are shown in Figs. \ref{fig:zvzacl} and \ref{fig:zvzapl}, and results
for the local scalar and pseudoscalar currents are show in in Figs.
\ref{fig:zpscl} and \ref{fig:zpspl}. The qualitative features of fattening
are the same for both actions: for thin link
actions all $z$'s are large in  magnitude and the differences between
``chiral partners'' (such as $z_V$ and $z_A$) are also large.
Either HYP action has tiny $z$'s (order unity) with differences an order
of magnitude smaller.

\begin{figure}[thb]
\begin{center}
\epsfxsize=0.7 \hsize
\epsffile{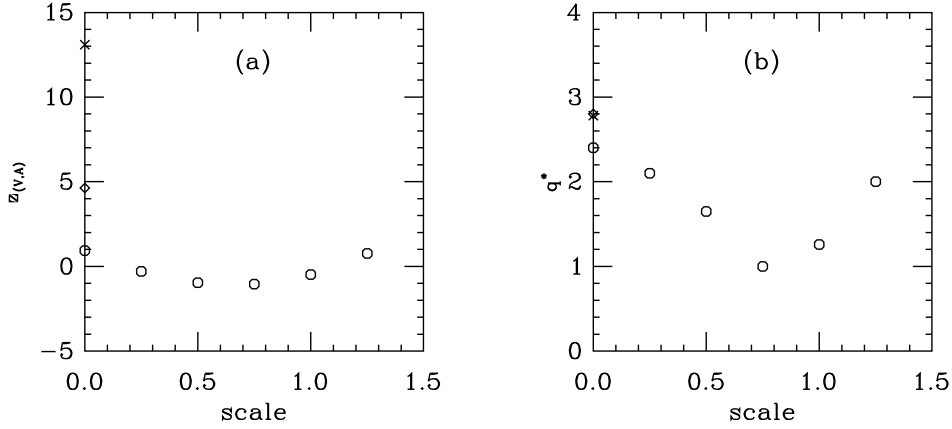}
\end{center}
\caption{
Coefficient of $g^2C_F/(16\pi^2)$ (a) and momentum
scale $q^*$ (b) for the local  vector (and axial vector) currents for 
overlap fermions with the planar action kernel and HYP links, at $x_0=1.6$.
Octagons label the planar action; also shown are the thin link Wilson
(cross) and clover (diamond) actions.
}
\label{fig:zvzaovplhyp}
\end{figure}

\begin{figure}[thb]
\begin{center}
\epsfxsize=0.7 \hsize
\epsffile{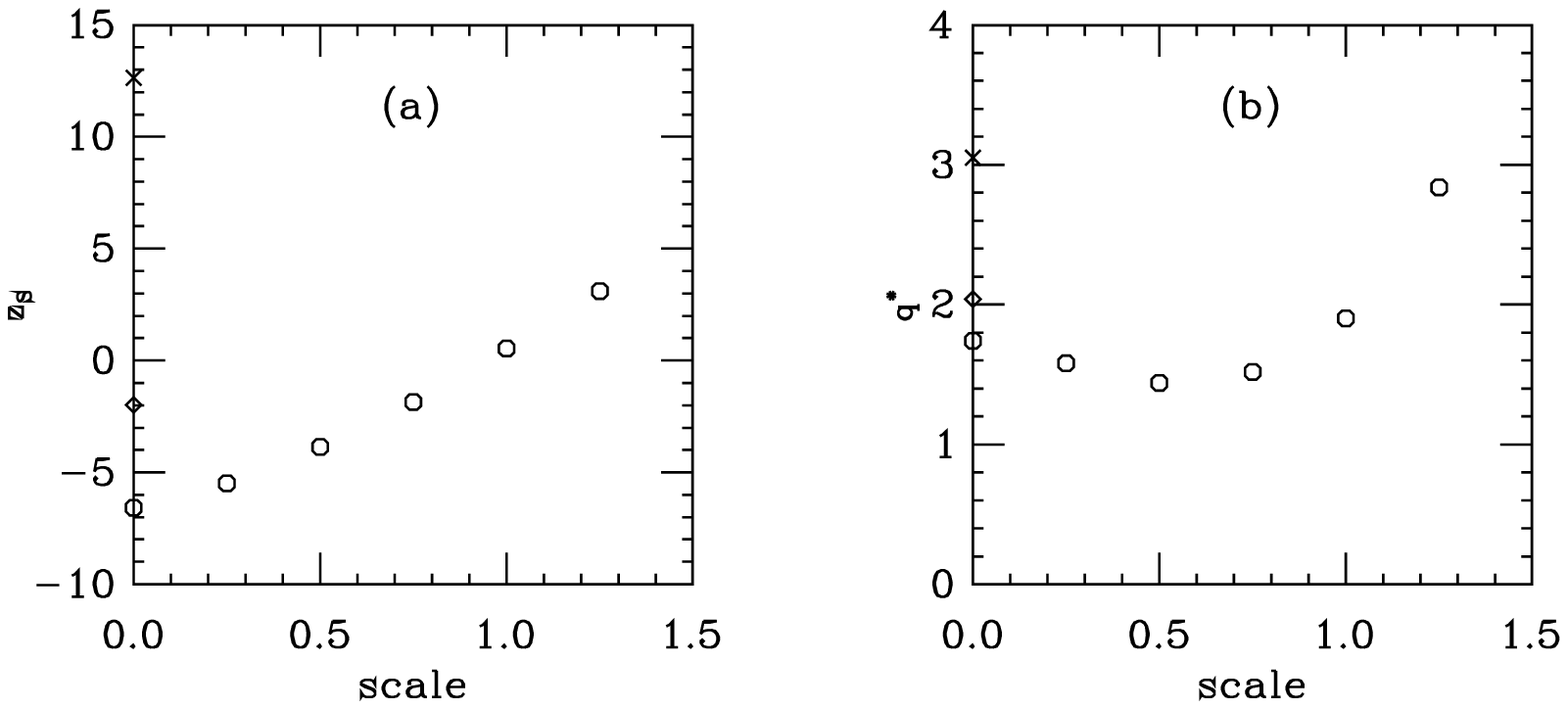}
\end{center}
\caption{
Coefficient of $g^2C_F/(16\pi^2)$ (a) and momentum
scale $q^*$ for the local scalar (and pseudoscalar) currents for 
overlap fermions with the planar action kernel and HYP links, at $x_0=1.6$.
Octagons label the planar action; also shown are the thin link Wilson
(cross) and clover (diamond) actions.
}
\label{fig:zpsovplhyp}
\end{figure}

Results for overlap actions parallel those for nonchiral actions:
matching factors drop when the clover term is included, and drop more
when the links are fattened. To illustrate this, I present results for
the  thin link Wilson and clover overlap, both with $x_0=1.6$,
and for the HYP planar overlap, in Figs. 
\ref{fig:zvzaovplhyp} and
\ref{fig:zpsovplhyp}.
A table of results for the HYP-planar overlap is given in 
Table \ref{tab:hypplov}.

\subsection{Four fermion operators}

Some sample results for four fermion operators
are shown in Table \ref{tab:bk}.
My results for  (ordinary non-overlap) Wilson fermions agree with the
DRED(EZ) results of Ref. \cite{ref:BDS} and the NDR results of
Ref. \cite{ref:GBS}.
My Wilson overlap results
 satisfy the connection
between $Z_+$, $Z_{V,A}$, and $Z_{P,S}$ of Ref. \cite{ref:GBS} and
Table \ref{tab:fourf}. For the special case of radius  $x_0=1$ they
agree with a calculation of P. Weisz\cite{ref:PWBK}.
They  differ by an overall additive
 factor of 14/3 (the precise value comes from Weisz; I have
 only determined this factor numerically)
 from the results of
Ref. \cite{Capitani:2000da}. NDR four fermion matching coefficients
for the Wilson overlap action for many radii
can readily be constructed from any desired operator using the
tables of bilinears from Ref. \cite{Alexandrou:2000kj} and
the results of Ref. \cite{ref:GBS} or
Table \ref{tab:fourf}  (though finding the $q^*$ scale will require actually 
doing some integrals).

\begin{figure}[thb]
\begin{center}
\epsfxsize=0.7 \hsize
\epsffile{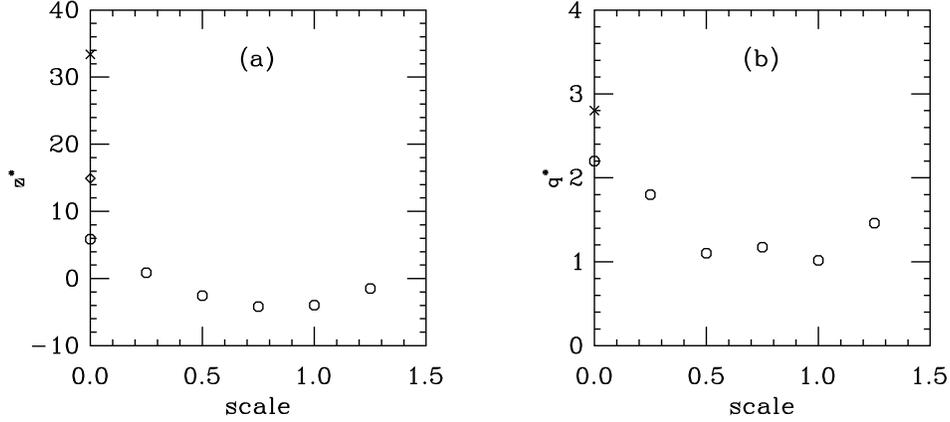}
\end{center}
\caption{
(a) Coefficient of $g^2/(16\pi^2)$ at scale $\mu a=1$  and (b) momentum
scale $q^*$ for $Z_+$ for matching lattice and NDR
overlap fermions with the planar action kernel, with $x_0=1.6$.
Octagons label the planar action; also shown are the thin link Wilson
(cross) and clover (diamond) actions.
}
\label{fig:ffermovplhyp}
\end{figure}

In Fig. \ref{fig:ffermovplhyp}
I show results for the HYP planar action kernel,
 as well as for the thin link Wilson and clover overlaps. In all cases I set
$x_0=1.6$. There is a large reduction in $z_+$ by converting from 
a Wilson kernel to a clover kernel even without fat links (recall that
the planar action includes a clover term). Fattening the links
further reduces $z_+$.

\begin{figure}[thb]
\begin{center}
\epsfxsize=0.7 \hsize
\epsffile{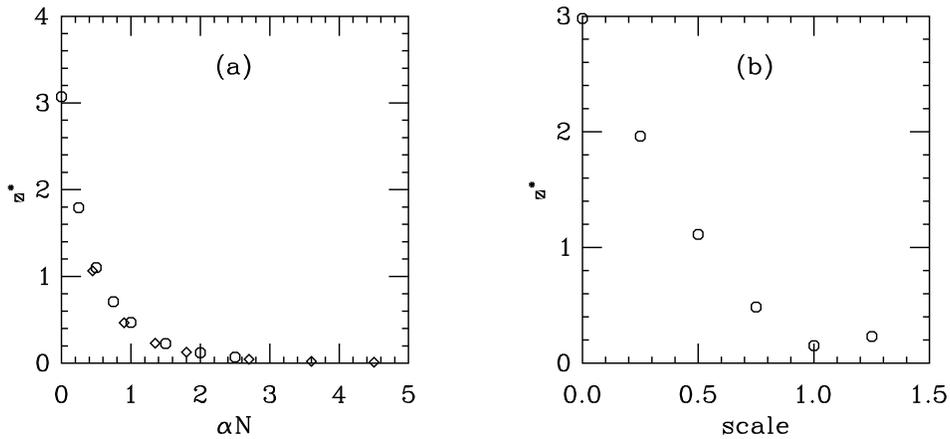}
\end{center}
\caption{
$Z^*$, the wrong-chirality mixing amplitude, for clover fermions
with (a) APE blocking (octagons for $\alpha=0.25$ and diamonds
for $\alpha=0.45$)
(b) HYP blocking, with the optimum parameters scaled by the shown overall 
scale factor.
}
\label{fig:zstar}
\end{figure}

Finally, to return to non-chiral actions, we can ask how fattening 
alters the mixing into different chiral sectors. This quantity
is parameterized by the coefficient $Z^* = -8K_1$.
From Eq. \ref{eq:vadesire},
$Z^* = z_V - z_A = 2(z_P-z_S)$. As we have already seen,
fattening pushes all the $z$'s closer to zero, and so their differences
also become small.
 Values of $Z^*$ for
the clover action with APE blocking and HYP blocking are shown in Fig.
\ref{fig:zstar}. 
As expected, either smearing can cut $Z^*$ by
over an order of magnitude.

The Wilson thin link fermion action value
$Z^*=9.6$ \cite{ref:BDS} is much greater
than even the thin-link clover result.
Converting to fat links without also turning on the clover term helps, but
will not be productive: for example, with scale factor unity, the
HYP-blocked Wilson action has $Z^*=2.29$, while including a $c_{SW}=1$
clover term cuts this number to 0.16.

Results for the operators $O_7$ and $O_8$
for the Wilson and planar overlaps  are shown in Table \ref{tab:b78}.

Let's use these results to
consider a numerical example of the ambiguities which will
afflict our calculation of mixing. In all cases we will match the lattice
and NDR calculations at a scale $q_i$ and run the NDR result to $q_f=2$ GeV
using the two-loop evolution equation. We will suppose we are using
the Wilson gauge action a coupling  at $\beta=5.9$ and
assume that the
 inverse lattice spacing is $1/a=1.58$ GeV. (These are
typical numbers from simulations \cite{ref:bkinprog}).
A standard
calculation a la Ref. \cite{ref:LM}
beginning with the logarithm of the average plaquette yields
$\alpha_{\overline {MS}}(qa=1)=0.199$ and
$\alpha_{\overline {MS}}(qa=\pi)=0.137$.
We begin with the HYP-planar overlap action and imagine just matching
at a scale $q_1=1/a$ or $\pi/a$, and running up. Then the full
matching and running matrix is
\bee
Z(q_f,q_i)Z(q_i) = \left(\matrix{ 0.979 & -0.006 \cr
                                             -0.039 & 1.099\cr}\right)
\ee
and the result at $q_i=\pi/a$ is
\bee
Z(q_f,q_i)Z(q_i) = \left(\matrix{ 0.998 & -0.030 \cr
                                             -0.043 & 1.012\cr}\right)
\ee
Next, we compute a ``lowest order'' $q^*$ from Eq. \ref{eq:matLM}.
We have $q^*_1 a = 3.6$, $q^*_2 a = 0.68$,
and
\bee
Z(q_f,q_i)Z(q_i) = \left(\matrix{ 0.966 & -0.020 \cr
                                             -0.009 & 1.112\cr}\right)
\ee
Notice that in all cases the matrices are nearly diagonal.

In contrast, the first choice for the Wilson action (just set $q^*a=1)$)
gives
\bee
Z(q_f,q_i)Z(q_i) = \left(\matrix{ 1.558 & -0.017 \cr
                                             -0.033 & 1.640\cr}\right),
\ee
matching at $q^*a=\pi$ and running gives
\bee
Z(q_f,q_i)Z(q_i) = \left(\matrix{ 1.412 & -0.071 \cr
                                             -0.050 & 1.304\cr}\right),
\ee
while using Eq. (\ref{eq:matLM}), with  $q^* a$=
(3.0, 2.3), we have
\bee
Z(q_f,q_i)Z(q_i) = \left(\matrix{ 1.459 & -0.016 \cr
                                             -0.042 & 1.530\cr}\right).
\ee
The off diagonal terms in $Z$ are small and not too dependent on the
running scheme, but the
diagonal coupling of the operators suffers a giant
renormalization. One probably should treat perturbation theory
results for the Wilson overlap cautiously for these factors.

\section{Conclusions}
Fattening the gauge connections of a lattice fermion action is a simple way
to reduce the size of perturbative matching coefficients.  HYP blocking
is a particularly felicitous choice: it combines large scale
smoothing with locality. It is clear from the results presented that
the qualitative features of fattening do not depend on the specific
choice of parameters. In situations where full chiral symmetry might
 not be necessary, the HYP-blocked clover action might be an attractive choice
for a light quark action.

In this work
I have only considered lattice actions with scalar and vector couplings and
nearest and next-nearest couplings. It would be easy to construct
the Feynman rules for ``hypercubic'' actions
(such as those of Refs. \cite{DeGrand:1999gp} or \cite{Bietenholz:2002ks}).
Techniques similar to the ones used here could be applied (with only a little
more effort) to the more complicated approximate fixed point or
overlap actions  used by several other authors,
whose kernels fill out the entire Clifford algebra\cite{morecomplicated}.

The result for $Z_+$ will be used in a lattice calculation of $B_K$
using the  HYP-blocked planar overlap \cite{ref:bkinprog}.
\section*{Acknowledgments}
I would like to thank Anna Hasenfratz, Roland Hoffman, and 
Francesco Knechtil for many discussions about HYP blocking,
and Weonjong Lee, Steve Sharpe, and Peter Weisz for correspondence.
I am very much indebted to Claude Bernard for introducing me  to 
lattice perturbation theory.
This work was supported by the
U.~S. Department of Energy with grant
DE-FG03-95ER40894.

\appendix
\section{Feynman Rules for  Planar-Action Fermions}
The non-overlap fermion action I am most interested in has
 scalar and vector
couplings to fermions offset on nearest-neighbor and 
diagonal-offset sites, and
minimal length gauge paths built of unitarized fat links connecting them.
The nearest-neighbor vector and scalar couplings are labeled
$\rho_1\def\rho_\mu^{(1)}$ and $\lambda_1$; the diagonal
($\vec r=\pm\hat\mu \pm\hat\nu$, $\nu\ne\mu$) couplings are $\rho_2$
 and $\lambda_2$. There is also a local scalar coupling $\lambda_0$.
For the Wilson or clover action, $\rho_2=\lambda_2=0$;
$\rho_1=\lambda_1= -1/2$.
For the ``planar'' action of Ref. \cite{ref:TOM_OVER}
$\lambda_1=-0.170$, $\rho_\mu^{(1)}=-0.177$ and diagonal
neighbors ($\vec r=\pm\hat\mu \pm\hat\nu$, $\nu\ne\mu$;
$\lambda_2= -0.061$, $\rho_\mu^{(2)}= \rho_\nu^{(2)}= -0.0538$.
 The constraint
 $\lambda_= -8\lambda_1 -24 \lambda_2$
 enforces masslessness on the free spectrum,
and $-1 = 2 \rho_\mu^{(1)} + 12\rho_\mu^{(2)}$ normalizes
the action to $-\bar \psi i \gamma_\mu \partial_\mu \psi$ in
the naive continuum limit. 

The free fermion action is then
\bee
d(p) = \lambda_0 + 2\lambda_1\sum_\mu\cos p_\mu
+4\lambda_2\sum_{\mu,\nu<\mu}\cos q_\mu \cos q_\nu
+i \sum_\mu \gamma_\mu \sin q_\mu (2\rho_1 +4\rho_2\sum_{\nu \ne \mu}\cos q_\nu).
\ee

Nearest-neighbor connections basically contribute rescaled
versions of the usual Wilson-action Feynamn rules.
The diagonal-offset neighbor gauge connections are taken to be
an average  of the two length-two shortest paths
connecting the fermions (each of which is a product of unitarized HYP
links). Neglecting the form factor arising from fattening, the
 first order vertex $V^{(1)}_\mu$ is
\beea
V^{(1,)}_\mu = 2i&\gamma_\mu\{(\rho_1 
  +\rho_2 \sum_{\nu \ne \mu}(\cos p_1 +\cos p_2)_\nu) \cos({{(p_1+p_2)_\mu}\over 2} \}
\nonumber \\
 & + \sum_{\nu \ne \mu} \gamma_\nu(2i\rho_2\sin ({{(p_1+p_2)_\mu}\over 2})
(\sin p_{1\nu}+\sin p_{2\nu}) )  \nonumber \\
 & -2(\lambda_1 + \lambda_2 
 \sum_{\nu \ne\mu}(\cos p_1 +\cos p_2)_\nu )\sin({{(p_1+p_2)_\mu}\over 2}) \nonumber .\\
\eea

I also include a clover term in $V^{(1)}$:
\bee
V^{(1,cl)}_\mu = -{1\over 2} C_{SW} \gamma_\mu \sum_{\nu \ne \mu}\gamma_\nu
\sin k_\nu \cos {{k_\mu}\over 2}.
\ee
(My definition of $C_{SW}$ would be unconventional for
Wilson fermions if $\lambda_1 \ne 1/2$,
or $r \ne 1$ in usual usage.) The planar action has $C_{SW}=1.03$.

The expression for the four-point vertex is 
long, but in all
the calculations done here, I only need an expression for the vertex
at zero fermion momenta  and
for its first derivative with respect to (equal) fermion momenta, also
at zero momentum. In that limit,
\bee
V^{(2)}_{\mu\nu} = \delta_{\mu\nu}(-2\lambda_1 -12 \lambda_2)
+ (1-\delta_{\mu\nu})4\lambda_2\sin{{ k_\mu}\over{2}} \sin{{ k_\nu}\over{2}})
\ee
and its derivative at zero external fermion momenta is
\bee
{{\partial V_\mu^{(2)}} \over {\partial p_\nu}} = p_\nu \gamma_\nu
(\delta_{\mu\nu}(-2\rho_1-24\rho_2)
+ (1-\delta_{\mu\nu})4\rho_2\sin {{k_\mu}\over{2}}\sin{{ k_\nu}\over{2}})
\ee

\section{Explicit formulas for fat links}
APE blocking: 
 The link after $n+1$ smearings is
related to the link after $n$ smearings by
\beea
V^{(n+1)}_\mu(x) = &
Proj_{SU(3)}((1-\alpha)V^{(n)}_\mu(x) \nonumber  \\
& +   \alpha/6 \sum_{\nu \ne \mu}
(V^{(n)}_\nu(x)V^{(n)}_\mu(x+\hat \nu)V^{(n)}_\nu(x+\hat \mu)^\dagger
\nonumber  \\
& +  V^{(n)}_\nu(x- \hat \nu)^\dagger
 V^{(n)}_\mu(x- \hat \nu)V^{(n)}_\nu(x - \hat \nu +\hat \mu) )).
\label{APE}
\eea
$V^{(n+1)}_\mu(x)$ is projected back onto $SU(3)$ after each step, and
 $V^{(0)}_\mu(n)=U_\mu(n)$ is the original link variable.
The momentum-space smearing factor for one level of smearing is
\begin{equation}
\tilde h_{\mu\nu}(q) = f(q)(\delta_{\mu\nu} - {\hat q_\mu\hat q_\nu
\over \hat q^2})
+ {\hat q_\mu\hat q_\nu\over \hat q^2}\ ,
\label{eq:APEBLOCK}
\end{equation}
with $\hat q_\mu ={2\over a} \sin({aq_\mu\over2})$ and
$f(q) =  1 - {\alpha\over6}\hat q^2$.
After $N$ smearings, $\tilde h_{\mu\nu}(q)$ becomes
$\tilde h^N_{\mu\nu}(q)$, which is just $\tilde h_{\mu\nu}$
with $f$  replaced by $f^N$. 

HYP blocking:
The momentum space version of HYP blocking is
\begin{eqnarray*}
B_\mu(q)&=&\sum_\nu h_{\mu\nu}(q)A_\nu(q)+ O(A^2),\ \ \textrm{where }\\
 h_{\mu\nu}(q)&=&\delta_{\mu\nu}\Big[1\!-\!\frac{\alpha_1}6
\sum_\rho \hat
q_\rho^2\Omega_{\mu\rho}(q)\Big]\!+\!\frac{\alpha_1}6\hat q_\mu
\hat q_\nu\Omega_{\mu\nu}(q),\\
\Omega_{\mu\nu}(q)&=&1+\alpha_2(1\!+\!\alpha_3)-\frac{\alpha_2}4
(1\!+\!2\alpha_3)(\hat q^2\!-\!\hat q_\mu^2\!-\!\hat
q_\nu^2) +\frac{\alpha_2\alpha_3}4{\prod_{\eta\neq\mu,\nu} \hat
 q_\eta^2}.\\
\end{eqnarray*}
with $\alpha_1=0.75$, $\alpha_2=0.6$, and $\alpha_3=0.3$
the favored parameterization of Ref. \cite{Hasenfratz:2001hp}.

\begin{table}
\begin{tabular}{cccc}
\hline
    &    $I_{latt}$  &  A   & B \\
$Z_V$    &  $4K_2 +2 K_1 + K_0 -S_1$    &   0   &   0  \\
$Z_A$    & $4K_2 -2 K_1 + K_0 -S_1$     &   0   &  0   \\
$Z_P$    &  $16K_2 -4K_1 +K_0 -S_1$    &   3   & -1/2    \\
$Z_S$    &  $ 16K_2 +4K_1 +K_0 -S_1$   &   3    &  -1/2   \\
\hline
\end{tabular}
\caption{ Ingredients for
matching coefficients for some local
operators.
}
\label{tab:vec}
\end{table}

\begin{table}
\begin{tabular}{cccc}
\hline
    &    $I_{latt}$  &  A   & B \\
$Z_+$    &  $8/3(K_2+K_0-S_1)$    &   -2    &   -5/3  \\
$Z_-$    & $80/3K_2 - 8/3(K_0-S_1)$     &   4    &  10/3   \\
$Z_{77}$    &  $4/3( 5K_2 + 2(K_0-S_1))$    &   -1    & 7/6    \\
$Z_{88}$    &  $4/3(32 K_2 + 2(K_0-S_1))$    &    8   &  -1/3   \\
$Z_{78}$ &  $12K_2$    &   3    &  -7/2   \\
$Z_{87}$ &  0    &    0   &  -3   \\
\hline
\end{tabular}
\caption{Ingredients for the (non-penguin parts) of
matching coefficients for some four-fermion
operators.
}
\label{tab:fourf}
\end{table}

\begin{table}
\begin{tabular}{|c|l|l||l|}
\hline
action &  process & $z_i$ & $q^*$ \\
\hline
thin planar   & & & \\
   & $S_0$ & 3.71     & 3.19    \\
   & $Z_V$ & -21.5     &2.8    \\
   & $Z_A$ &  -20.0   &  2.8  \\
   & $Z_S$ & -26.2    & 2.4   \\
   & $Z_P$ & -29.3    & 2.4   \\
\hline
HYP planar & & & \\
   & $S_0$ & 0.26     & 2.37    \\
   & $Z_V$ & -2.27    & 1.55   \\
   & $Z_A$ & -2.19    & 1.54   \\
   & $Z_S$ & -0.92    & 1.50   \\
   & $Z_P$ & -1.11    & 1.51   \\
\hline
\end{tabular}
\caption{Table of Z-factors and $q^*$'s for planar and HYP-planar
actions, defined so $Z_i = 1 + z_i g^2 C_f /(16\pi^2)$.
$C_{SW}=1.03$.
}
\label{tab:hyppl}
\end{table}
\begin{table}
\begin{tabular}{|c|l|l|l|}
\hline
action &  process & $z_i$ & $q^*$ \\
\hline
thin & & & \\
  &  $Z_{V,A}$ & -14.8    & 3.27   \\
  &  $Z_{P,S}$ & -39.2    & 2.79   \\
\hline
HYP & & & \\
  &  $Z_{V,A}$ & -0.945    & 2.36   \\
  &  $Z_{P,S}$ & -0.592    & 2.05   \\
\hline
\end{tabular}
\caption{Table of Z-factors and $q^*$'s for  thin link and
 HYP link naive fermion actions.
}
\label{tab:naive}
\end{table}

\begin{table}
\begin{tabular}{|c|l|l|}
\hline
 process & z & $q^*$ \\
\hline
  $Z_{V,A}$ & -0.489    & 1.26   \\
   $Z_{P,S}$ & 0.53    & 1.96   \\
\hline
\end{tabular}
\caption{Table of Z-factors and $q^*$'s for bilinears,
 HYP-planar overlap action.
}
\label{tab:hypplov}
\end{table}

\begin{table}
\begin{tabular}{|c|l|l|}
\hline
action &  $z_+$ &   $q^*$   \\
\hline 
thin Wilson             & 33.4 & 2.86  \\
thin Clover          & 14.3 & 4.07  \\
$0.45\times 10$ APE Clover      & -5.8 & 1.05 \\   
$0.45\times 10$ APE  planar & -5.67 &0.95  \\
HYP planar              & -3.97 & 0.92  \\
\hline
\end{tabular}
\caption{Table of Z-factors into NDR and $q^*$'s for $O_+=O_1+O_2$ for
overlap actions with various kernels. In all cases $x_0=1.6$,
$C_{SW}=1$ for clover action and 1.03 for planar action.
 $Z_+ = 1 + \alpha_s(q^*)/(4\pi)(z_+ + 4 \log(a\mu))$.
}
\label{tab:bk}
\end{table}

\begin{table}
\begin{tabular}{|c|l|l|l|l|}
\hline
action & ${\cal Z}_{77}$ &${\cal Z}_{78}$ & ${\cal Z}_{87}$ & ${\cal Z}_{88}$\\
\hline 
thin Wilson ${\cal Z}$ & 36.29 & -3.51 & -3 & 34.85  \\
thin Wilson ${\cal L}_1$ & 70.57 & 1.56 & -0.003 & 75.26  \\
thin Wilson ${\cal L}_2$ & 17.6 & -4.64 & 9.84 & 19.32 \\
\hline 
HYP planar  ${\cal Z}$& -0.65 & -1.98 & -3 & 2.41   \\
HYP planar  ${\cal L}_1$& 1.52 & 2.72 &-0.003 & 9.69   \\
HYP planar  ${\cal L}_2$& 4.94 & -1.65 & 9.84 & 27.50  \\
\hline
\end{tabular}
\caption{Table of Z-factors and $q^*$'s for $O_7$ and $O_8$ for
overlap actions with various kernels. In all cases $x_0=1.6$,
$C_{SW}=1.03$ for the planar action.
 $Z_{ij} = \delta_{ij} + \alpha_s(q^*)/(4\pi){\cal Z}_{ij}$.  
${\cal L}_1$ and ${\cal L}_2$ are the integrals of ${\cal Z}$,
weighted by $\log ~q^2$ and $\log^2 ~q^2$, respectively. 
}
\label{tab:b78}
\end{table}


\end{document}